\documentclass[usenatbib]{mnras}
\usepackage{graphicx}
\usepackage[space]{grffile}
\usepackage{latexsym}
\usepackage{textcomp}
\usepackage{longtable}
\usepackage{multirow,booktabs}
\usepackage{amsfonts,amsmath,amssymb}
\usepackage{natbib}
\usepackage{url}
\usepackage{hyperref}
\usepackage[utf8]{inputenc}
\usepackage[english]{babel}
\usepackage{times}

\title[The Mass Distribution of Molecular Clouds in M83]{The Varying Mass Distribution of Molecular Clouds Across M83}

\author[Freeman et al.]{Pamela Freeman$^{1}$, Erik Rosolowsky$^{1}$,  J.~M.~Diederik Kruijssen$^{2,3}$,
\newauthor Nate Bastian$^{4}$, and Angela Adamo$^{5}$ \\
$^{1}$Dept. of Physics, University of Alberta, Edmonton, AB, Canada\\
$^{2}$Astronomisches Rechen-Institut, Zentrum f\"{u}r Astronomie der Universit\"{a}t Heidelberg, M\"{o}nchhofstra\ss e 12-14, 69120 Heidelberg, Germany\\
$^{3}$Max-Planck Institut f\"{u}r Astronomie, K\"{o}nigstuhl 17, 69117 Heidelberg, Germany\\
$^{4}$Astrophysics Research Institute, Liverpool John Moores University, Liverpool, UK\\
$^{5}$Department of Astronomy, Stockholm University, Stockholm, Sweden \\}

\begin{document}

\date{Accepted 2017 February 24}

\pagerange{\pageref{firstpage}--\pageref{lastpage}} \pubyear{2002}

\maketitle

\label{firstpage}

\begin{abstract}
The work of \citet{Adamo_2015} showed that the mass distributions of young massive stellar clusters were truncated above a maximum-mass scale in the nearby galaxy M83 and that this truncation mass varies with galactocentric radius.  Here, we present a cloud-based analysis of ALMA CO($1\to 0$) observations of M83 to search for such a truncation mass in the molecular cloud population.  We identify a population of 873 molecular clouds in M83 that is largely similar to those found in the Milky Way and Local Group galaxies, though clouds in the centre of the galaxy show high surface densities and enhanced turbulence, as is common for clouds in high-density nuclear environments.  Like the young massive clusters, we find a maximum-mass scale for the molecular clouds that decreases radially in the galaxy.  We find the most massive young massive cluster tracks the most massive molecular cloud with the cluster mass being $10^{-2}$ times that of the most massive molecular cloud.  Outside the nuclear region of M83 ($R_{g}>0.5$ kpc), there is no evidence for changing internal conditions in the population of molecular clouds, with the average internal pressures, densities, and free-fall times remaining constant for the cloud population over the galaxy.  This result is consistent with the bound cluster formation efficiency depending only on the large-scale properties of the ISM, rather than the internal conditions of individual clouds.
\end{abstract}%


\bibliographystyle{mnras}

\section{Introduction}

Molecular gas is the host of all known star formation in the local and distant Universe.  The average properties of the star formation process point to roughly constant star formation rate per unit free fall time \citep[e.g.,][]{Krumholz_2007}. However, there is emerging evidence,  particularly in dense gas environments, of variations in the molecular gas depletion timescale ($\tau_{\mathrm{dep}} \equiv \Sigma_{\mathrm{H2}}/\dot{\Sigma}_{\star}$, where $\Sigma_{\mathrm{H2}}$ is the molecular gas surface density and $\dot{\Sigma}_{\star}$ is the star formation rate surface density).  Recent work in the local 
\citep{Longmore_2013, Kruijssen_GalCen,Usero_2015,Bigiel_2015,Pereira_Santaella_2016,Bigiel_2016} and high-redshift \citep{Genzel_2015,Aravena_2016,Scoville_2016} Universe point to significant variation in depletion times, converging to the sense that higher density environments have shorter depletion times.  Whether the changes in depletion time reflect a different mode of star formation or a variation along a continuum \citep[e.g.,][]{Krumholz_2011} remains unsettled.  

Star formation is often parameterized as uniform mass rate of star production, but the organization of the resulting stellar structures also shows significant evolution with star formation rate, and by correlation, with the molecular gas density.  In particular, the fraction of stars formed in bound clusters \citep[$\Gamma$,][]{Bastian_2008} is seen to correlate with kpc-scale gas surface density  and weakly with the star formation rate \citep{Larsen_2002,AdamoBastian_2015, Kruijssen_2016}.  The origin of this correlation has been attributed to differences in the structure of the molecular (star forming) interstellar medium \citep[ISM,][]{Kruijssen_2012}.  This is based on the idea that there is a correlation between the structure of the star forming ISM and the resulting clusters that form out of that gas.  In particular, the upper end of the cluster mass function appears to be truncated at a mass scale \citep{Larsen_2009,Bastian_2011,Konstantopoulos_2013, Diederik_Kruijssen_2014} and that truncation mass changes with galactic environment \citep{Diederik_Kruijssen_2014,Adamo_2015}.  

When the molecular ISM is partitioned into molecular clouds, the mass distribution of the population also shows a characteristic truncation mass \citep{Williams_1997,Rosolowsky_2005a} and the truncation mass also varies with the  changing properties of the galactic environment \citep{Rosolowsky_2007,  Colombo_2014,Hughes_2015}.  Despite preliminary evidence that these two truncation masses are linked \citep{Diederik_Kruijssen_2014}, this correlation has not yet been demonstrated for a homogeneous sample of molecular clouds and stellar clusters.  Furthermore, the maximum-mass scales of the molecular clouds and clusters have not been well linked back to the cloud formation process, though models of cloud formation should predict the resulting characteristic mass \citep[e.g.,][]{Duarte_Cabral_2016,Pan_2016}.  Several different cloud formation scenarios have been proposed \citep{Dobbs_2014} and the evolving maximum-mass scale provides a clear observational approach for evaluating those formation mechanisms.

The nearby galaxy M83 provides an excellent opportunity to evaluate the evolving mass distribution of molecular clouds in conjunction with the changing cluster properties.  As the nearest \citep[$D=4.5$~Mpc;][]{Thim_2003}, nearly face-on ($i = 24^{\circ}$), massive spiral galaxy \citep[$M_{\star} = 6.4\times 10^{10}~M_{\odot}$;][]{Lundgren_2004b}, M83 is an obvious target for exploring molecular cloud properties.  Archival Hubble Space Telescope data have already been analyzed, showing a significant change in both the fraction of star formation that results in bound clusters, and the changing truncation masses of young massive cluster populations \citep{Silva_Villa_2013,Adamo_2015}.  This latter work found that the cluster mass distribution followed a Schechter function, i.e. a power-law mass distribution with an index of $-2$ and an exponential cutoff above a truncation mass.  We therefore investigate whether the molecular cloud population, which must serve as the progenitors of these clusters, follows a similar mass distribution with truncation.  To explore this question, we utilize archival data from the Atacama Millimeter/Submillimeter Array (ALMA), which can provide exceptional imaging data of this nearby galaxy.  In particular, we use the high-quality $^{12}$CO($1\to0$) data set observed as part of ALMA project 2012.1.00762.S (PI: A. Hirota). 

In this paper, we present a cloud-based analysis of the molecular emission in M83, searching for environmental variation in the cloud populations.  In Section \ref{sec:data}, we present the ALMA data and report on the relevant limitations that affect our analysis.  In Section \ref{sec:cprops}, we use the {\sc cprops} algorithm \citep{Rosolowsky_2006} to decompose the molecular emission into the population of molecular clouds.  We compare the properties of these clouds to the populations seen in the Milky Way and nearby galaxies.  For comparison to the studies of clusters, we also analyze the mass distributions of the clouds.  In Section \ref{sec:discussion}, we interpret the results of the analysis in terms of theoretical interpretations and the empirical results derived from cluster analysis.   

\section{ALMA Data}
\label{sec:data}
This project uses observations made by ALMA under project 2012.1.00762.S as proposed by Hirota et al.  We used the Quality Assurance, step 2 data (QA2) downloaded from the Japanese Virtual Observatory site, which images the CO($1\to0$) line at $1.34\times 0.83$ arcsecond resolution, corresponding to $29\times 18$ pc at the 4.5 Mpc distance of M83, with a geometric mean of 22.8 pc.  The imaged data cube has a velocity resolution of 2.57 km~s$^{-1}$.  With a median brightness sensitivity of $\sigma=0.89$~K per beam, the data are well suited for the identification and decomposition of GMCs \citep{Rosolowsky_2006}.  Through this analysis, we adopt a CO-to-H$_2$ conversion factor of $X_{\mathrm{CO}} = 2.0\times 10^{20}$~(K~km~s$^{-1}$)$^{-1}$~cm$^{-2}$ \citep{Bolatto_2013}.  With this conversion factor, the data cube has a $1\sigma$ mass surface density sensitivity of 9.9 $M_{\odot}~\mathrm{pc}^{-2}$ and a per-beam mass sensitivity of $6.0\times 10^3~M_{\odot}$.   

Since the QA2 data delivered as part of the project do not include total power or short spacing data, the resulting image is affected by some negative sidelobes near bright sources from missing spatial information in the interferometer maps.  These artifacts are most noticeable toward the bright emission in the nucleus of the galaxy.  The data are also likely missing flux from large-scale, diffuse CO emission \citep[e.g.,][]{Pety_2013}.  However, filtering out this diffuse component facilitates the identification of the compact, star-forming molecular clouds.  Apart from the limitations of interferometer-only imaging, the quality of the data is excellent and shows no signs of calibration artifacts.    The data are thus well-suited for the task at hand, namely identifying the GMCs in the galaxy as bright compact features in the CO emission and then characterizing their properties.

\begin{figure*}
\begin{center}
\includegraphics[width=\textwidth]{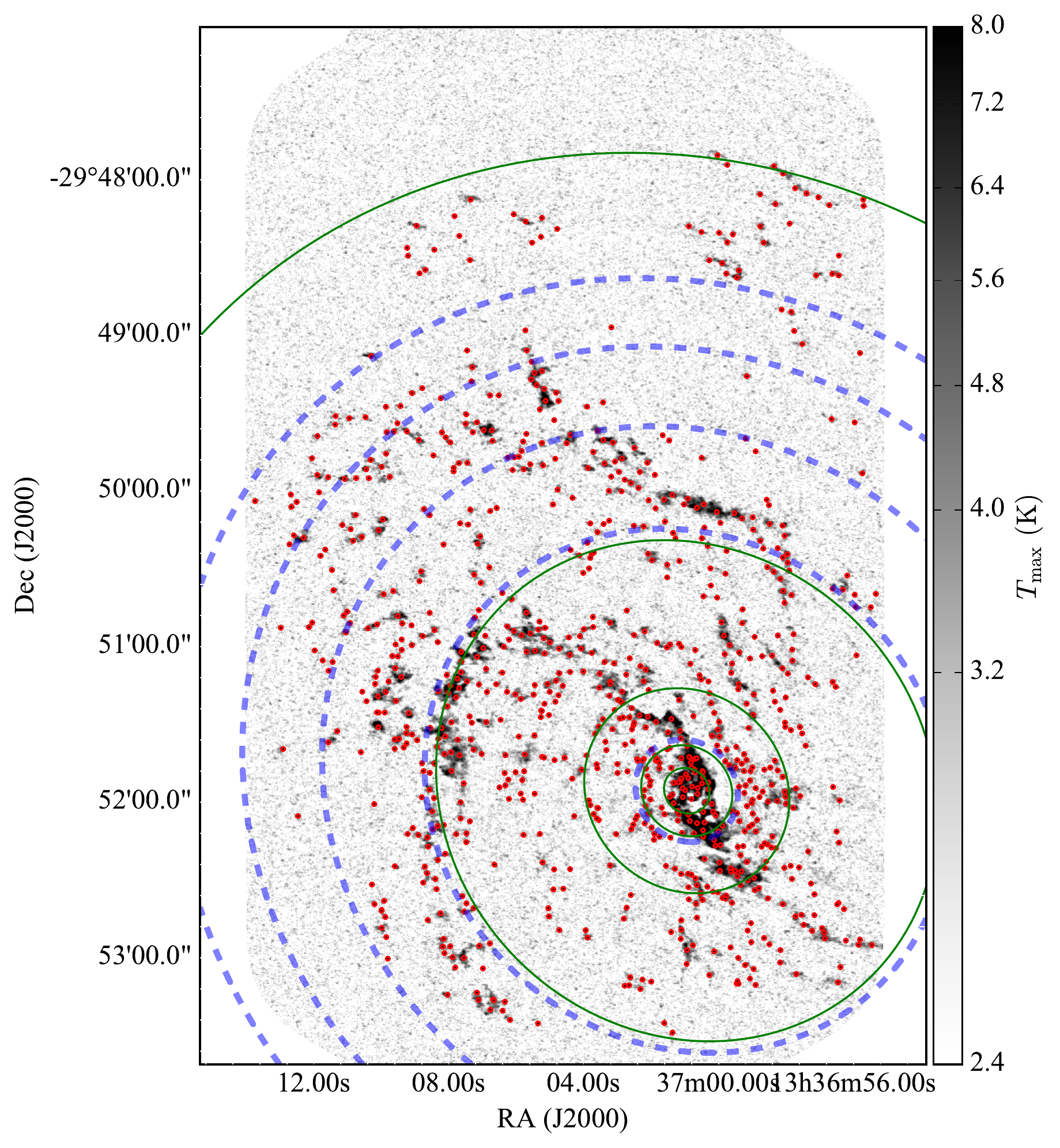}
\caption{Maximum antenna temperature map of M83 derived from the ALMA CO(1-0) data.  The location of clouds identified by the {\sc cprops} algorithm and included in the catalog are indicated as red points. The blue, dashed contours indicate the radial bins in the galaxy used in the equal-area mass distribution analysis in Section \ref{sec:mspec}, following the boundaries established in \citet{Adamo_2015}.  The green, solid contours show radial bins used in the equal-mass analysis, where each bin has an equal molecular mass (CO luminosity) \label{fig:tmaxmap}.%
}
\end{center}
\end{figure*}

\section{Analysis}
\label{sec:cprops}
\subsection{GMC properties}
We characterize the properties of the extracted clouds in the CO data to assess whether we are identifying clouds that can be compared to GMCs seen in the Milky Way, or whether the emission structures in the M83 data are better described as Giant Molecular Associations (GMAs), which are larger scale structures of molecular gas \citep{Rand_1990}.

We identify molecular clouds in the CO emission line data using the {\sc cprops} algorithm \citep{Rosolowsky_2006}\footnote{We use the {\sc cpropstoo} implementation at \url{http://github.com/akleroy/cpropstoo}}.  
We utilize their recommended algorithm for identifying GMCs in interferometer data described as follows.  The algorithm first calculates a spatially varying estimate of the noise in the map by calculating the rms ($\sigma(\alpha,\delta)$) of signal-free channels.  Emission is then identified as those pixels in the (three-dimensional) data cube that are larger than $2.5\sigma(\alpha,\delta)$ in two adjacent velocity channels.  This emission mask is then extended to include all connected pixels which are larger than $1\sigma(\alpha,\delta)$ in two adjacent channels.  We test the masking algorithm by applying these criteria to the data set scaled by $-1$.  We find no false positives are included so the masking criteria are likely robust.  

The masked emission is then divided into individual molecular clouds using a seeded watershed algorithm, with individual clouds being defined by local maxima (the ``seeds'') that are separated by at least 40 pc spatially or 5.14 km~s$^{-1}$ in velocity.  Any pair of local maxima in the same contiguous region of the mask is also required to be at least $2.5\sigma(\alpha,\delta)$ above the saddle point of emission connecting those maxima.  The watershed algorithm then assigns a given pixel of emission to the local maximum that is connected to the pixel by the shortest path contained entirely within the emission mask.  This approach extracts 873 clouds.

\begin{figure}
\includegraphics[scale=0.8]{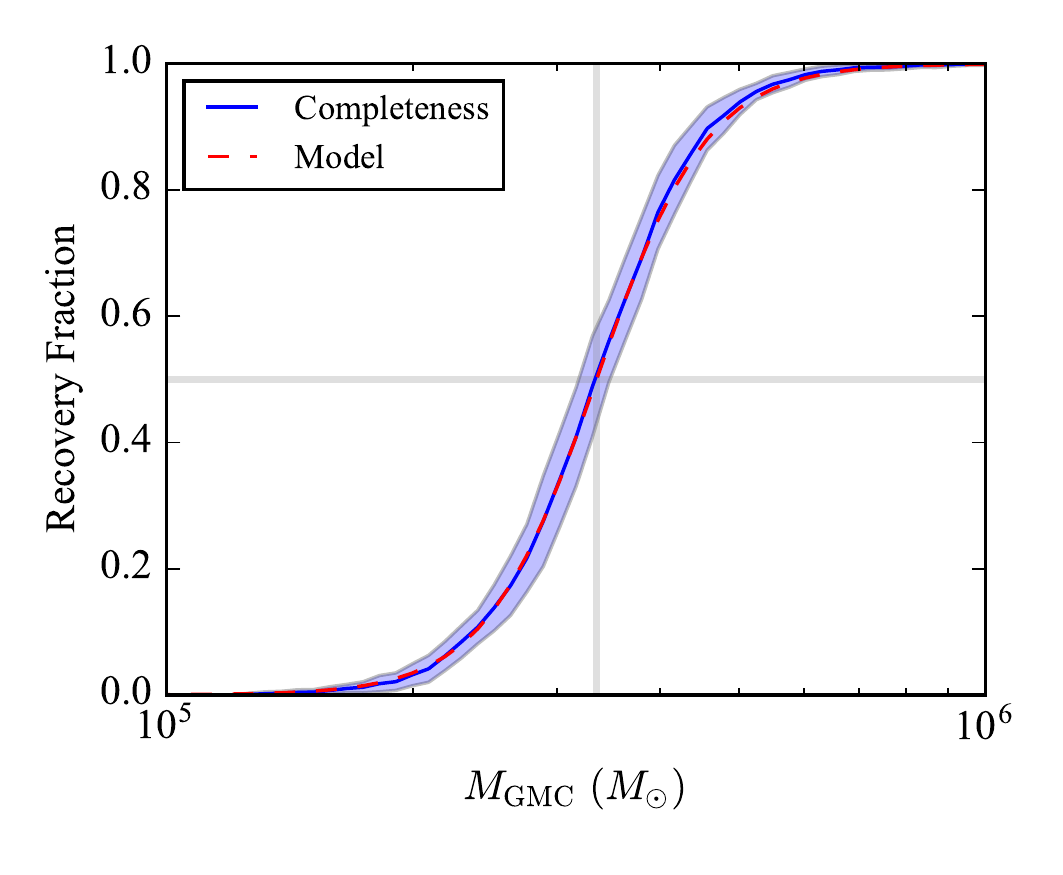}
\caption{\label{compfrac} Fraction of GMCs recovered by the cloud identification algorithm in a false source injection test.  The solid blue line indicates the mean over 10 separate trials of 350 clouds at a given input cloud masses but different locations in the data set. The shaded area indicates the standard deviation around that fraction.  We use this analysis to estimate a completeness limit of 50\% for GMC masses with $M>3.35\times 10^5~M_{\odot}$.  The red dashed line indicates the model of the completeness fraction used to correct the source counts.} 
\end{figure}

\begin{figure*}
\begin{center}
\includegraphics[width=1\textwidth]{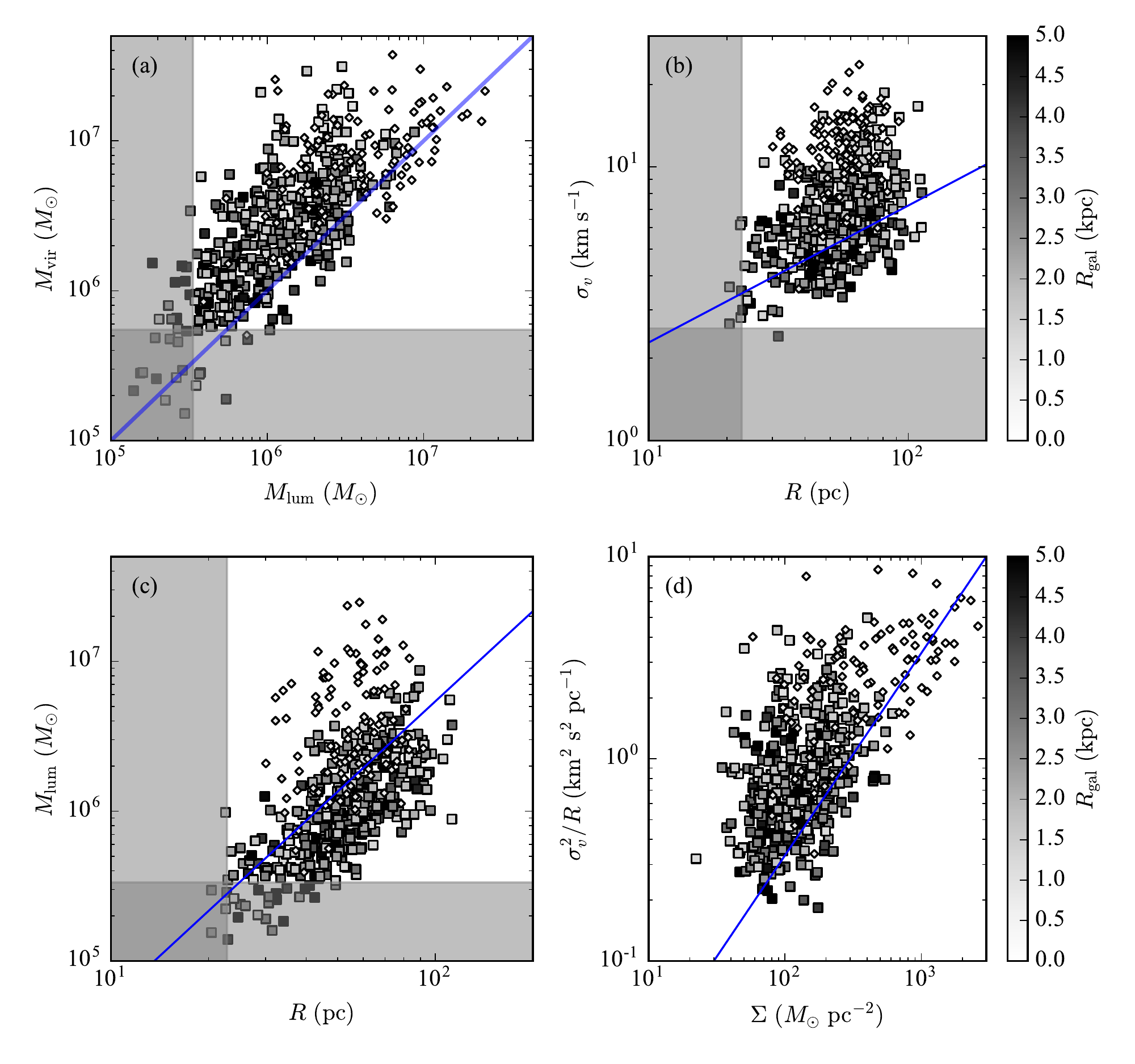}
\caption{\label{larson_figure} Correlations between the properties of molecular clouds in M83.  The solid lines show relationships seen in the Milky Way clouds from the catalog of \citet{Solomon_1987}.  The shaded regions indicate where spatial (22.8 pc), spectral (2.57 km s$^{-1}$) and sensitivity limits ($M_\mathrm{lum}>3.35\times 10^5~M_{\odot}$) will censor data.  (a) This panel shows the correlation between the virial and luminous mass estimates for the clouds, with the shade of the plotting symbol representing the distance from the centre of the galaxy.  Clouds with galactocentric radius $R_{g}<1\mbox{ kpc}$ are plotted with diamonds and clouds in the remainder of the galaxy are indicated with squares.  The two mass estimates correlate well, particularly at the high mass end, where the signal-to-noise ratio is highest. (b) The size-line width relationship for clouds in M83 shows good agreement with the relationship measured in the Milky Way (solid line).  However, the clouds in the centre of M83 are displaced above the relationship.  (c) The mass-radius relationship shows most clouds have surface densities close to that of the Milky Way, and clouds in the centre of M83 show significantly higher surface densities. (d) The correlation between surface density and velocity dispersion on a 1 pc scale (here represented by the square of that quantity, $\sigma_v^2/R$) shows that most clouds agree well with the locus of self-gravitation, which is a corollary of the agreement between the luminous and dynamical masses shown in panel (a).
}
\end{center}
\end{figure*}

Since the primary goal of this study is to investigate the mass distributions, we establish a completeness limit through false source injection into a signal-free portion of the ALMA data set.  The fake sources are three-dimensional Gaussian clouds of a specified mass with luminosities, sizes, and line widths set by the characteristic relationships established in the Local Group population \citep{Solomon_1987, Bolatto_2008}.  Our subsequent analysis (e.g., Figure \ref{larson_figure}) shows this is a good approximation to the M83 GMC population.  We inject a set of 350 sources on a fixed grid into the data set, identify and characterize the sources with the same source selection parameters as used in the real source identification and measure the fraction of the sources recovered and their recovered properties.  This process is repeated for 51 different mass values distributed logarithmically between $10^5~M_{\odot}$ and $10^6~M_{\odot}$.  This analysis is repeated 10 times for different, random spatial offsets for the grid of fake sources.  Figure \ref{compfrac} summarizes the source fraction recovery and is the basis for establishing a 50\% completeness limit of $M_{\mathrm{GMC}} > 3.35\times 10^5~M_{\odot}$.  We model the recovery fraction [$f(M)$] as a logistic function and fit the parameters of the function to obtain an analytic expression for the survey completeness:
\begin{equation}
f(M) = \left\{1 + \exp\left[-14.8\left(\log_{10} \frac{M}{M_{\odot}}-5.525\right)\right]\right\}^{-1}
\label{eqn:compmodel}
\end{equation}
We use this expression to correct the source counts in the survey in fitting mass distributions.  Of the 873 clouds identified, 711 are above the 50\% completeness limit.

We determine the macroscopic properties of the GMCs in the system by calculating moments of the emission line data.  To account for emission below the edge of the emission mask, we correct each measured property by a Gaussian correction factor following \citet{Rosolowsky_2005}.  This correction is necessary to avoid bias in low signal-to-noise data \citep{Rosolowsky_2006}.  The correction factor is a function of the ratio of the peak emission in the GMC to the value of the emission at the edge of the emission mask.  The factor is calculated assuming the cloud has a Gaussian profile and estimates the corrections to be applied to the moments measure over the emission mask relative to the moments that would be measured for infinite sensitivity.  We select this correction method over others based on it recovering the properties of clouds in our fake source injection test.  The extrapolation method discussed in \citet{Rosolowsky_2006} leads to mass values that are biased low by 25\%.  Since the clouds are only marginally resolved by a Gaussian restoring beam, their emission profiles will be close to Gaussian.

We calculate the CO luminosity of the molecular clouds ($L_{\mathrm{CO}}$) by integrating the emission associated with each cloud, with a Gaussian correction.  The luminous mass is calculated by scaling by a single CO-to-H$_2$ conversion factor:
\begin{equation}
M_{\mathrm{lum}} = \alpha_{\mathrm{CO}} L_{\mathrm{CO}},
\end{equation}
where $\alpha_{\mathrm{CO}}=4.35~M_{\odot}\mbox{ pc}^{-2}~\mathrm{(K~km~s^{-1})^{-1}}$.  The velocity dispersion is the Gaussian-corrected, emission-weighted second moment of the velocity axis, corrected for the channel width.  Similarly radius of the cloud is the root-mean-square of the Gaussian-corrected, emission-weighted second moments of the major and minor spatial axis of the emission.  The radius is also corrected for the instrumental response by assuming an elliptical beam and subtracting its width in quadrature \citep[for details, see ][]{Rosolowsky_2006}.  The algorithm corrects for the case where the beam and cloud position angles are not aligned. 

The virial mass of the cloud is calculated from the radius and line width of the molecular cloud: $M_{\mathrm{vir}} = 5 \sigma^2 R/G$.  Comparing the virial and luminous masses gives insight to the dynamical nature of the molecular clouds identified in the data.  The average surface density is calculated from the luminous mass: $\Sigma = M_{\mathrm{lum}}/(\pi R^2)$.  Typical uncertainties are assessed by the bootstrapping method in {\sc cprops} and are 0.2 dex in the velocity dispersion and line width and 0.3 dex in the mass estimates (both virial and luminous), though these errors grow when the signal-to-noise ratio approaches the $2.5\sigma$ threshold.

We only include clouds in our final analysis with $M>3.35\times 10^{5}~M_{\odot}$, corresponding formally to a $\sim 50\sigma$ aggregate detection of a molecular cloud.  This is a factor of $\sim 8$ larger than the minimum mass that would be admitted by our masking procedure, but represents a conservative treatment of the spatial filtering artifacts in the centre of the galaxy, which make cloud identification less certain.

Correlations of these macroscopic properties give clues to the nature of the molecular medium.  We compare the properties of the molecular clouds to those seen in the Milky Way study of \citet{Solomon_1987} because that work measured GMC properties using similar techniques as we do here.  In Figure \ref{larson_figure},  we correlate the GMC properties and compare the result to the trends seen in the S87 data.  First, we see (panel a) that there is good agreement between the virial and luminous masses in these clouds, and this is seen throughout the system.  We shade each datum by the galactocentric radius to highlight the variation in cloud properties across the face of the disk.  The most massive clouds are found in the centre of the galaxy, but these extreme clouds still show good agreement between the two mass estimates.  Both mass estimates will be subject to $\sim 0.3~\mathrm{ dex}$ uncertainties, but even in high quality data, there remains about 0.5 dex of intrinsic scatter \citep[see also][]{Heyer_2009}. 

\begin{figure*}
\begin{center}
\includegraphics[width=1\textwidth]{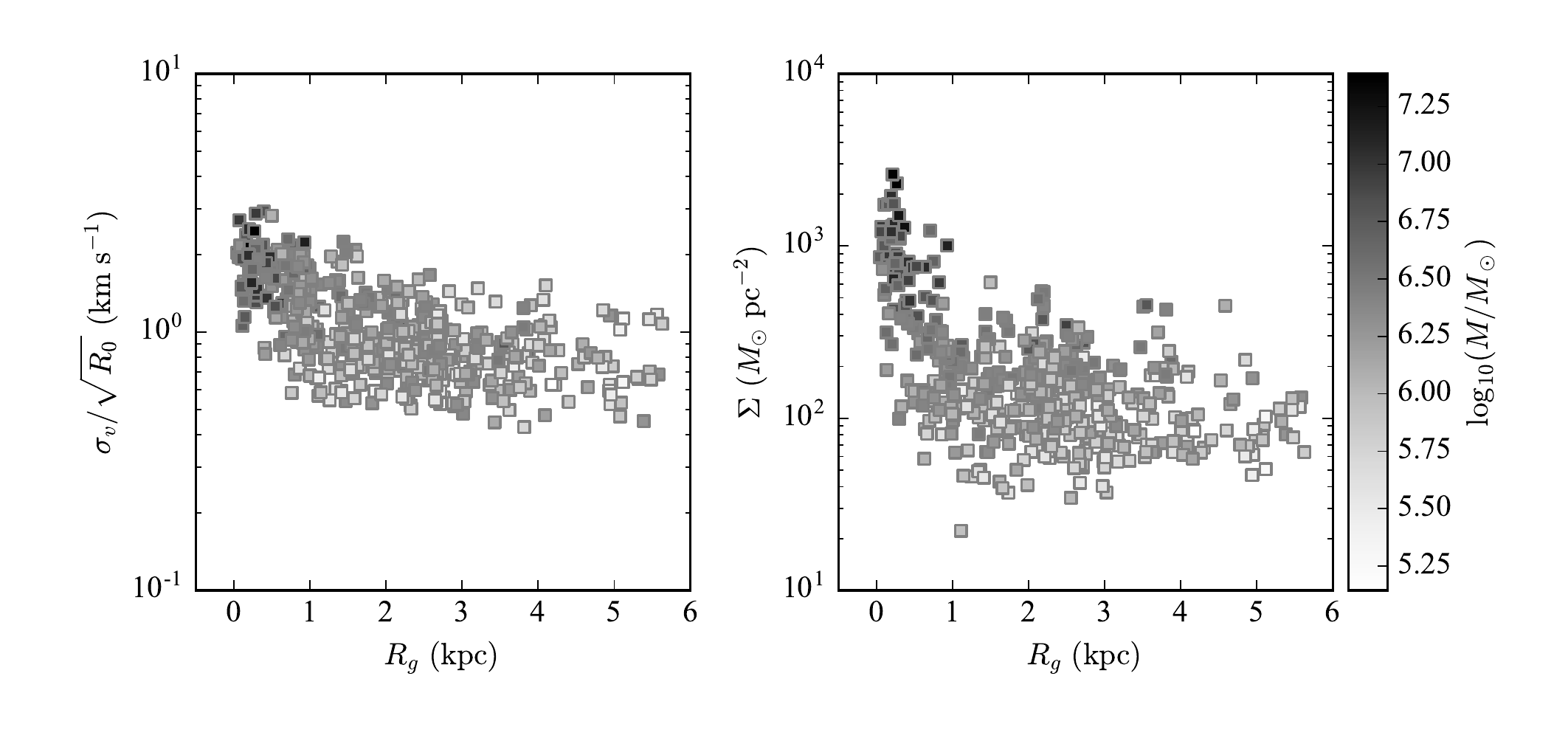}
\caption{Changing internal properties of molecular clouds with radius in the galaxy.  The left panel illustrates the changing turbulent line width normalized to a 1 pc scale [$R_0 \equiv (R/1~\mathrm{pc})$], assuming a size-line width relationship with an index of 0.5.  Clouds in the centre of the galaxy are more turbulent than clouds in the outer part of the galaxy.  The right panel illustrates the changing surface density of clouds in the galaxy with clouds in the centre of M83 having higher surface density than clouds in the outskirts.  The greyscale represents the mass of the clouds.  \label{fig:radial}%
}
\end{center}
\end{figure*}

The radius-velocity dispersion plot (Figure \ref{larson_figure}b) shows the size-line width scalings in these clouds.  The Milky Way relationship shows a good lower bound for the population, but there is significant scatter to higher line widths at a given cloud radius.  These offset clouds are found in the centre of the galaxy and are also associated with the higher mass clouds.  Such objects are typically seen in molecule rich environments, where the surface densities of clouds increase significantly \citep{Oka_2001,Rosolowsky_2005,Heyer_2009,Hughes_2013, Leroy_2015}.  Such clouds are also seen as outliers in the mass-radius plot (Figure \ref{larson_figure}c).  Clouds at galactocentric radius $R_{g}>0.5\mbox{ kpc}$ have a median surface density of $\langle \Sigma \rangle = 170~M_{\odot}\mbox{ pc}^{-2}$ but clouds inside this radius have a median surface density of $\langle \Sigma \rangle = 700~M_{\odot}\mbox{ pc}^{-2}$.  For self-gravitating clouds, the internal gas pressures will have $P_{\mathrm{int}} \propto \Sigma^2$, so the clouds in the centre of the galaxy show markedly higher internal pressures than disk clouds (see also Figure \ref{fig:radial}).

Given the other changes across the face of the galaxy, the clouds all appear to show gravitational binding energies comparable to their kinetic energies.  Figure \ref{larson_figure}d shows the correlation between surface density and the turbulent line width on 1 pc scales $\sigma_0 = \sigma_v/R^{1/2}$ (the $y$-axis shows the square of this quantity).  \citet{Heyer_2009} noted that these quantities correlate even in clouds that show line widths and surface densities that depart significantly from the Milky Way relations.  This relationship is thus roughly equivalent to the relationship plotted in Figure \ref{larson_figure}a.

In Figure \ref{fig:radial} we plot the internal conditions within GMCs as a function of galactocentric radius to highlight the trends seen in Figure \ref{larson_figure}.  The two panels show that clouds in the central region are more turbulent on a fixed (1 pc) scale and have higher surface densities than clouds in the disk of the galaxy.  For $R_g\gtrsim 1$~kpc, the clouds seem to have constant values for these measurements of internal conditions with significant scatter.  The increased turbulence and surface densities are typical of clouds in molecule-rich environments such as the centre of the Milky Way or high-redshift galaxies \citep{Oka_2001, kruijssen13}.   Even so, the clouds in the centre of M83 retain a good balance between gravitational and kinetic energies (Figure \ref{larson_figure}a).  Overall, the GMCs found in M83 are consistent with the results seen in other galaxies given recent studies \citep[e.g.,][]{Hughes_2010, Hughes_2013, DonovanMeyer_2012, DonovanMeyer_2013, rebolledo_2015}.

\subsection{Mass distributions}
\label{sec:mspec}

\begin{figure*}
\begin{center}
\includegraphics[width=1\textwidth]{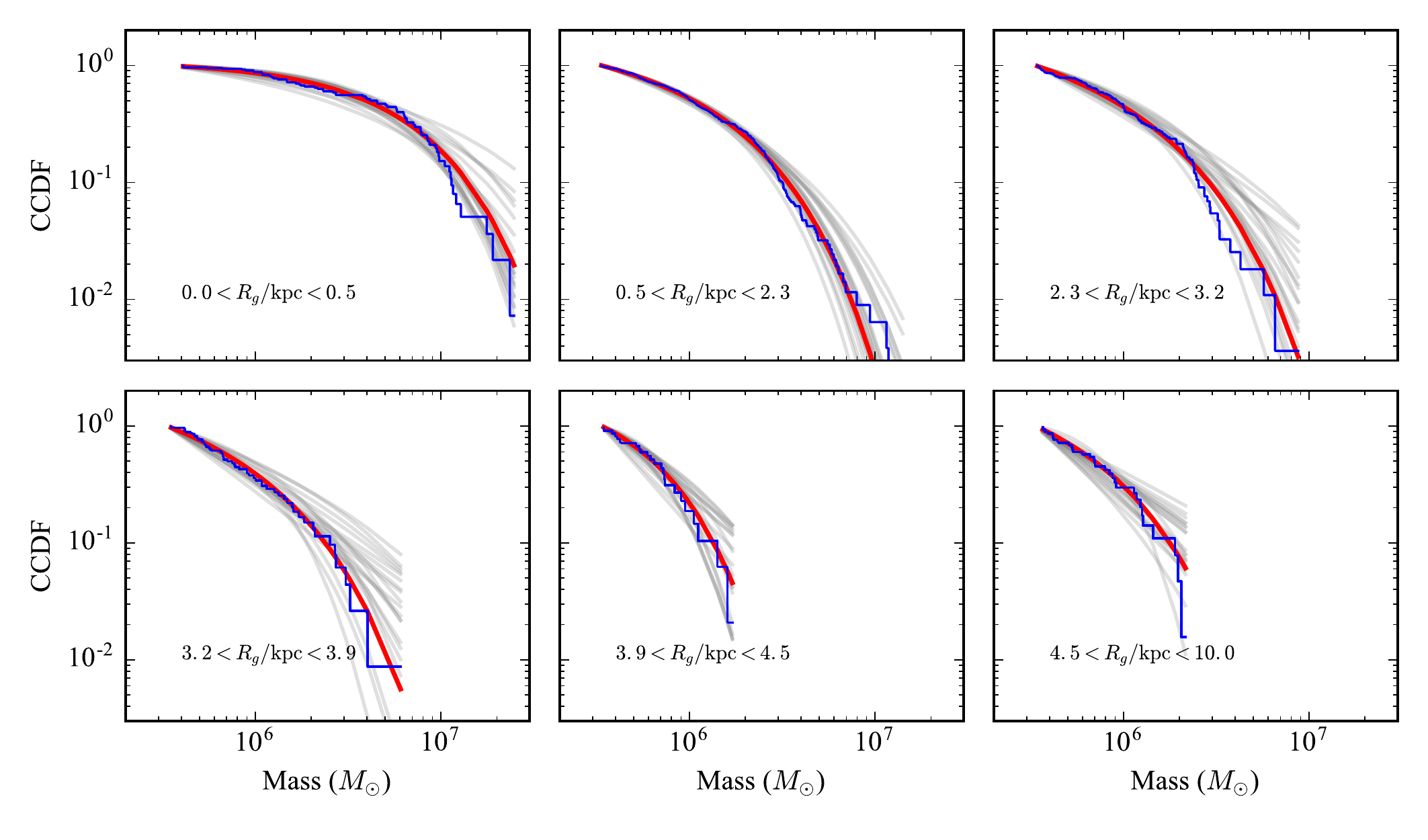}
\caption{Fits to the complementary cumulative mass distribution functions for the same (equal-area) regions used in the \citet{Adamo_2015} work.  The empirical, completeness-corrected distributions are shown as the blue curve.  The optimized truncated power-law fit is shown as a continuous red curve.  The light grey curves show 20 draws from the MCMC posterior parameter distributions indicating the range of fits that are consistent with the empirical CDF under the Anderson-Darling goodness-of-fit measure.  For the inner three radial bins there is good evidence for a characteristic truncation mass.  In the outer three bins, the evidence for a truncation is not as strong (see Table \ref{table:properties}), but the maximum-mass clouds seen in these bins is appreciably lower than in the inner bins.
\label{fig:massfits}%
}
\end{center}
\end{figure*}

\begin{figure}
\begin{center}
\includegraphics[width=\columnwidth]{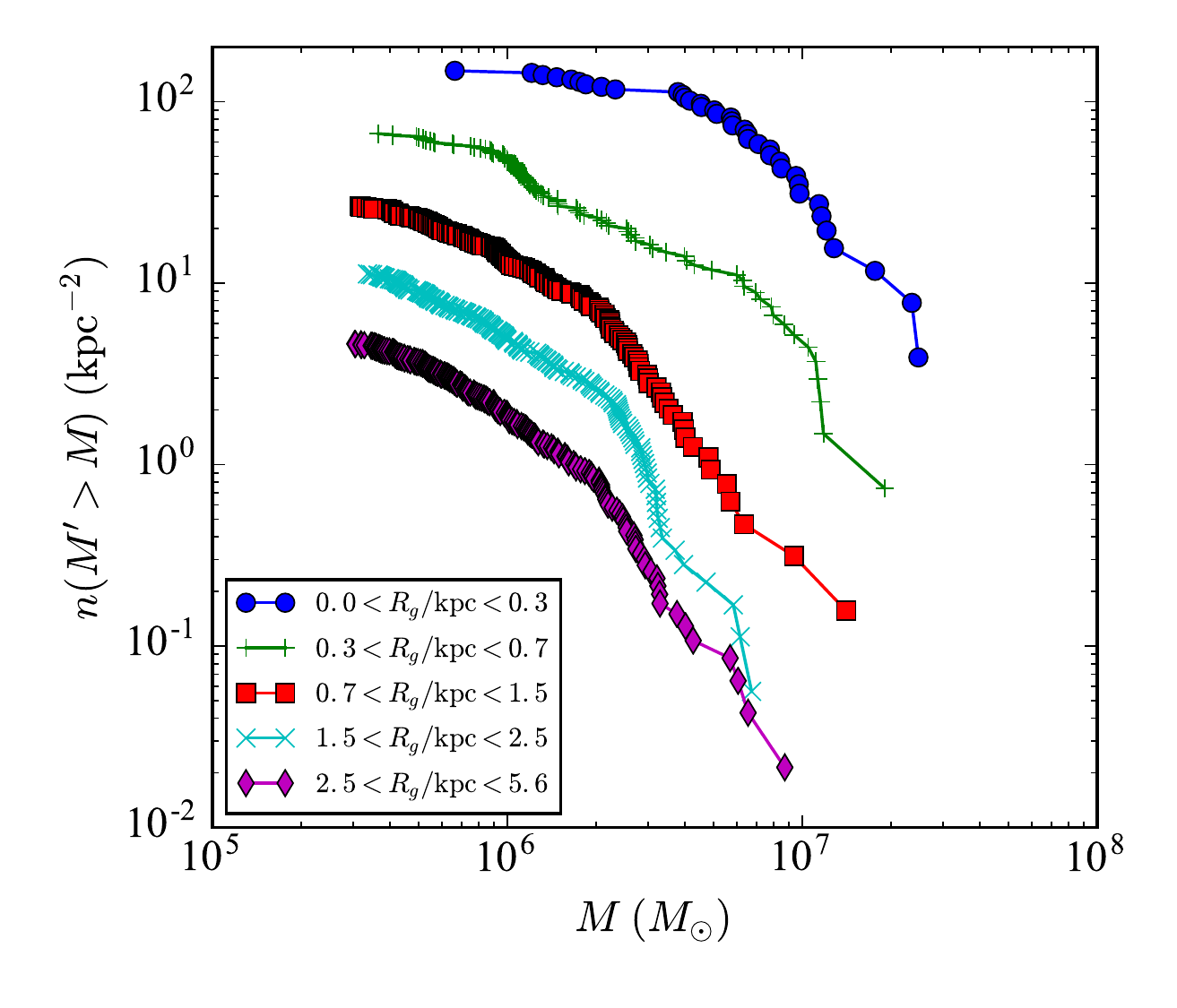}
\caption{Cumulative mass distributions, divided into five regions with equal molecular gas mass ($2.5\times 10^8~M_{\odot}$).  The equal-area binning shows similar behaviour as the by-area binning (Figure \ref{fig:massfits}, namely more massive clouds being found in the centre of the galaxy).  This binning emphasizes the differences between the cloud population. In particular, several of the mass distributions show truncations at the high mass end, and the bar region $1.5<R_g/\mathrm{kpc}<2.5$ has a deficit in high mass clouds compared to the clouds both inside and outside this radius.  \label{fig:massdist}%
}
\end{center}
\end{figure}

Figures \ref{fig:massfits} and \ref{fig:massdist} plot the cumulative mass distribution functions for five radial bins in the galaxy, where the bins have roughly equal area in Figure \ref{fig:massfits} and equal total mass in Figure \ref{fig:massdist}.  The normalization to equal area bins adopts the binning used in \citet{Adamo_2015} to facilitate direct comparison with that work.  That binning divides the galaxy into regions with equal area in the original optical data.  The data set we analyze here does not span the full range of angles around the galaxy, though the areas are still significant fractions of this range (see Figure \ref{fig:tmaxmap}).  The mass distributions clearly change over the face of the galaxy and several of the bins show some evidence for truncation at high masses.  

Figure \ref{fig:massdist} emphasizes the changing mass distributions.  Each distribution there has been binned radially into groups with equal total mass ($M_{\mathrm{tot}}=1.4\times 10^{8}~M_{\odot}$).  This binning emphasizes the radial changes in how the same amount of molecular gas mass is distributed in each annulus, with the distribution in the centre of the M83 having significantly higher mass clouds than the outer disk, though cloud blending will affect this result (see Section \ref{sec:systematics}).  Of note, the mass distribution of clouds between $1.5\mathrm{~kpc}<R_{g}<2.5\mathrm{~kpc}$ shows a truncation at the high mass end that is not seen as clearly in the other regions.  A similar truncation was seen in the nuclear ring region of M51 by \citet{Colombo_2014}, where the lack of high mass clouds was attributed to dynamical suppression of high mass cloud formation. Since this region includes most of the bar in M83, it is reasonable to expect that similar mechanisms are at work here.  The grouping into bins of equal mass is illustrative of the different mass distributions, but we focus on the equal-area binning for the remainder of the paper since it can be compared to the cluster mass distribution.

\begin{table*}
\begin{tabular}{r r r r r r r}
\hline
 & \multicolumn{5}{c}{Radial Bin (kpc)} \\
Property & 0$-$0.45 & 0.45$-$2.3 & 2.3$-$3.2 & 3.2$-$3.9 &  3.9$-$4.5 & $>4.5$ \\
\hline
Number of Clouds & 71 &  468 &  180 &   73 &   35 &   46 \\

Number of Clouds $> 3.35\times 10^{5}~M_{\odot}$ &   69 &  391 &  138 &   57 &   24 &   32 \\
$M_{\mathrm{max,GMC}}$ ($10^6 M_{\odot}$)&  24.7 & 14.1 &  8.7 &  6.1 &  1.7 &  2.2\\
$\langle M \rangle_5$ ($10^6 M_{\odot}$) &  17.8 & 10.0 &  4.9 &  3.2 &  1.3 &  1.8 \\
$\beta_{\mathrm{cluster}}$ &  $\cdots$ & $-1.9$ & $-2.2$ & $-2.2$ & $-2.7$ & $\cdots$\\
$M_{c, \mathrm{cluster}}$ ($M_{\odot}$)& $\cdots$ & 4.00 & 1.00 & 0.55 & 0.25 & $\cdots$ \\
$M_{\mathrm{max,cluster}}$ ($10^5 M_{\odot}$) & $\cdots$ & 1.5 & 2.0 & 0.6 & 0.3 & $\cdots$ \\
$M_{\mathrm{ALMA}}$ ($10^8 M_{\odot}$) & 3.7 & 6.0 & 2.2 & 0.7 & 0.4 & 0.4\\
$M_{\mathrm{SEST}}$ ($10^8 M_{\odot}$) & 1.6 & 10. & 3.2 & 1.8 & 1.1 & 1.7\\
\hline
$\beta_{\mathrm{PL}}$ &   $-1.3^{+ 0.1}_{- 0.1}$ & $-1.7^{+ 0.1}_{- 0.1}$ & $-1.8^{+ 0.1}_{- 0.1}$ & $-1.9^{+ 0.1}_{- 0.2}$ & $-2.2^{+ 0.2}_{- 0.4}$ & $-2.0^{+ 0.2}_{- 0.3}$\\
$M_{c,\mathrm{PL}}$ ($10^6 M_{\odot}$) & $\infty$& $\infty$ & $\infty$ & $\infty$ & $\infty$ & $\infty$  \\  
$\log_{10} p_{\mathrm{PL}}$ & $-6.04$ & $-10.64$ & $-2.92$ & $-1.33$ & $-0.58$ & $-0.62$\\ 
\hline
$\beta_{\mathrm{Sch}}$ &  $-2$ & $-2$ & $-2$ & $-2$ & $-2$ & $-2$ \\
$M_{c,\mathrm{Sch}}$ ($10^6 M_{\odot}$) & $\cdots$ & $\cdots$ & $\cdots$ & $\cdots$ &  $ 5.8^{+150.4}_{- 2.8}$ & $26.5^{+171.5}_{-20.4}$\\  
$\log_{10} p_{\mathrm{Sch}}$ &  $-\infty$ & $-\infty$ & $-\infty$ & $-\infty$ & $-0.47$ & $-0.57$ \\
\hline
$\beta_{\mathrm{TPL}}$ & $-0.2^{+ 0.2}_{- 0.5}$ & $-0.7^{+ 0.2}_{- 0.3}$ & $-1.0^{+ 0.2}_{- 0.7}$ & $-0.9^{+ 0.1}_{- 0.9}$ & $-0.1^{+ 0.5}_{- 2.2}$ & $-0.8^{+ 0.6}_{- 1.3}$ \\
$M_{c,\mathrm{TPL}}$ ($10^6 M_{\odot}$) & $ 7.0^{+10.5}_{- 1.8}$ & $ 2.1^{+ 1.4}_{- 0.4}$ & $ 2.2^{+30.9}_{- 0.6}$ & $ 1.7^{+66.5}_{- 0.2}$ & $ 0.5^{+64.8}_{- 0.1}$ & $ 1.0^{+50.5}_{- 0.4}$ \\
$\log_{10} p_{\mathrm{TPL}}$ & $-0.50$ & $-0.82$ & $-0.93$ & $-0.43$ & $-0.16$ & $-0.20$ \\
\hline
\end{tabular}
   \caption{\label{table:properties} The properties of GMCs and stellar clusters \citep[from][]{Adamo_2015} in M83, for bins of equal area. Values were derived for clouds more massive than $3.35\times 10^5 M_\odot$, and clusters more massive than 5000 M$_\odot$.   The maximum cloud mass $M_{\mathrm{max}}$ and the geometric mean of the five most massive clouds $\langle M\rangle_5$ show non-parametric measurements of the changing cloud mass distribution.  The total masses in each of the bins indicates the mass recovery, by comparing the total mass found in the ALMA catalog ($M_{\mathrm{ALMA}}$) to that calculated from the surface density profile of \citet[][ $M_{\mathrm{SEST}}$]{Lundgren_2004}. The cluster mass distributions are parameterized with a pure power-law index ($\beta_{\mathrm{cluster}}$), a Schechter mass cutoff ($M_{c,\mathrm{cluster}}$),  and a maximum mass in each bin ($M_{\mathrm{max,cluster}}$).  For each of the three models for the mass distribution considered in this work we report the power-law index $\beta$ and cutoff mass ($M_c$) based on the optimization described in the text.  We also report the log-likelihood $(\log_{10} p)$ given by the Anderson-Darling statistic for the optimized fit parameters.  The three models considered are a pure power-law (PL),  Schechter function (Sch) and a truncated power-law (TPL).}
\end{table*}

For each of the equal-area bins in the above analysis, we modeled the complementary cumulative mass distribution function (CCDF):
\begin{equation} 
\mathrm{CCDF} = 1-\frac{N(>M)}{N_{\mathrm{tot}}}=1-\frac{1}{N_{\mathrm{tot}}}\int^\infty_M \frac{dN}{dM'}dM'
\end{equation}
We consider three different models to the CCDF, where each model follows the general form of the mass distribution function of 
\begin{equation} 
\label{eq:powerlaw}
\frac{dN}{dM} = M^{\beta} \exp\left(-\frac{M}{M_c}\right),
\end{equation}
which is a power-law mass distribution with an exponential truncation.  The cutoff (truncation) mass in the distribution is $M_c$ and the index of the distribution is $\beta$.  The three models we consider are: (1) a {\it pure power-law} distribution, letting $M_c\to\infty$ in Equation \ref{eq:powerlaw} and allowing $\beta$ to vary; (2) a {\it Schechter function}, fixing $\beta=-2$ while letting $M_{c}$ vary; and (3) a {\it truncated power-law} where both $\beta$ and $M_c$ are free.  The pure power-law function has been considered in previous studies of the molecular cloud mass distributions \citep{Solomon_1987,Rosolowsky_2005a}.  The \citet{Schechter_1976} form of the mass distribution is expected for the gravitational fragmentation of a gas distribution below a characteristic fragmentation mass (e.g., the Jeans mass).   \citet{Adamo_2015} use the Schechter function in their analysis of young massive clusters, so we also consider it here.  In all cases, we limit the fits to clouds with masses greater than the 50\% completeness limit: $M>3.35\times 10^{5}~M_{\odot}$.  

To estimate the parameters of the mass distribution and their uncertainties, we build off the formalism developed by \citet{Clauset_2009}, which uses a maximum likelihood framework and a Kolmogorov-Smirnov test to assess goodness-of-fit.  Of note, this approach provides likelihoods for favouring one functional representation over another.  However, the approach does not provide estimates of parameter uncertainties, which precludes evaluating the significance of changes in the parameters.  To address this limitation, we refine this approach.  First, we use an Anderson-Darling measurement of the similarity between the observed cumulative distribution function (CDF) and the predicted value.  The Anderson-Darling statistic \citep{stephens1986} is given by 
\begin{equation}
A = n \int_{-\infty}^{\infty} \frac{[C_n(x) - C(x)]^2}{C(x)[1-C(x)]} dx,
\end{equation}
where $C_n(x)$ is the empirical CDF and $C(x)$ is the theoretical CDF.  In this analysis, we use a {\em completeness-corrected} cumulative distribution function for the cloud mass $M$:
\begin{equation}
C_n(M) = \frac{C_{n,\mathrm{obs}}(M)}{f(M)},
\end{equation}
where the observed cumulative distribution function, $C_{n,\mathrm{obs}}(M)$, is corrected by the observed completeness fraction, $f(M)$ given by equation \ref{eqn:compmodel}. The statistic $A$ is analogous to the distance $D$ in the Kolmogorov-Smirnov test.  There $D$ represents the maximum distance between empirical and theoretical CDFs.  Here $A$ is quadratic difference between the two CDFs with the denominator serving as a weighting function that emphasizes differences in the tails of the two distributions.  Like the Kolmogorov-Smirnov test, the Anderson-Darling statistic is frequently used to perform a test of whether the observed data are unlikely to be drawn from the theoretical CDF.  The Anderson-Darling test has been shown to have greater statistical power than the Kolmogorov-Smirnov and distinguishing between distributions \citep{Stephens1974}, provided the distribution functions are known.  

The Anderson-Darling test requires a full set of tabulated values for the test statistic that is set by the distribution function and number of data.  These cutoff values are usually drawn from Monte Carlo simulations \citep[e.g.,][]{choulakian2001}. Such a tabulation does not exist for all our functions or at our data points, so we must perform our own Monte Carlo simulations.  To assess the probability that a observed CDF is different from the theoretical CDF, we perform a series of Monte Carlo simulations.  We draw a number of random data from our three different distribution functions with known parameters ($N=30$ to $10^3$, spanning the sample sizes for which we are estimating parameters).  We then compute the statistic $A$ when comparing to the theoretical CDF of those same parameters.  This comparison accounts for the distribution of statistic values $A$ under many different finite draws from the probability density function.  Our results show that for $N>30$, we are in the limit where the test statistic $A$ is approximately distributed as $\exp(-A/(2n))$ for all our probability density functions.  Thus, the log-likelihood of a random draw from a known PDF generating a statistic of value $A$ is approximately $-A/(2n)$.  Following the approach of \citet{Clauset_2009}, we can then use this model of the log-likelihood to identify the parameters of the PDF that best represent the observed data.  

Since this approach yields an approximate distribution for the log-likelihood, we can also sample from this distribution using a Markov-Chain Monte Carlo sampler to provide credible distributions of distribution parameters that are not inconsistent with the observed data.  To do this sampling, we first optimize the parmeters of the distribution that minimize $A$ and then use the {\sc emcee} sampler \citep{emcee} to sample the log-likehood function over the parameters $\beta$ and $\log_{10} (M_{c}/M_{\mathrm{min}})$, where $M_{\min}=3.35\times 10^5~M_{\odot}$ is the completeness limit.  We assume ignorant priors:  $\beta\in [-20,20]$ and $\log_{10} (M_{c}/M_{\mathrm{min}}) \in [0, 3]$.  The upper limit of  $\log_{10} (M_{c}/M_{\mathrm{min}})=3$ is set to be a approximately one third of the total molecular mass in the observed area.  We use ten sampler chains initialized around the optimal fit, 200 steps of burn in and then 2000 steps of sampling, thinning to every 10th step based on the scale of the chain autocorrelation.  The quantiles of the posterior distributions of the parameters gives the credible ranges for the parameters.  We report asymmetric uncertainties in the parameters for the 16th and 84th percentiles of the parameter distributions, representing the $\pm 1 \sigma$ error bars that would be observed for a normal distribution of uncertainties.
\begin{figure}
\includegraphics[scale=0.7]{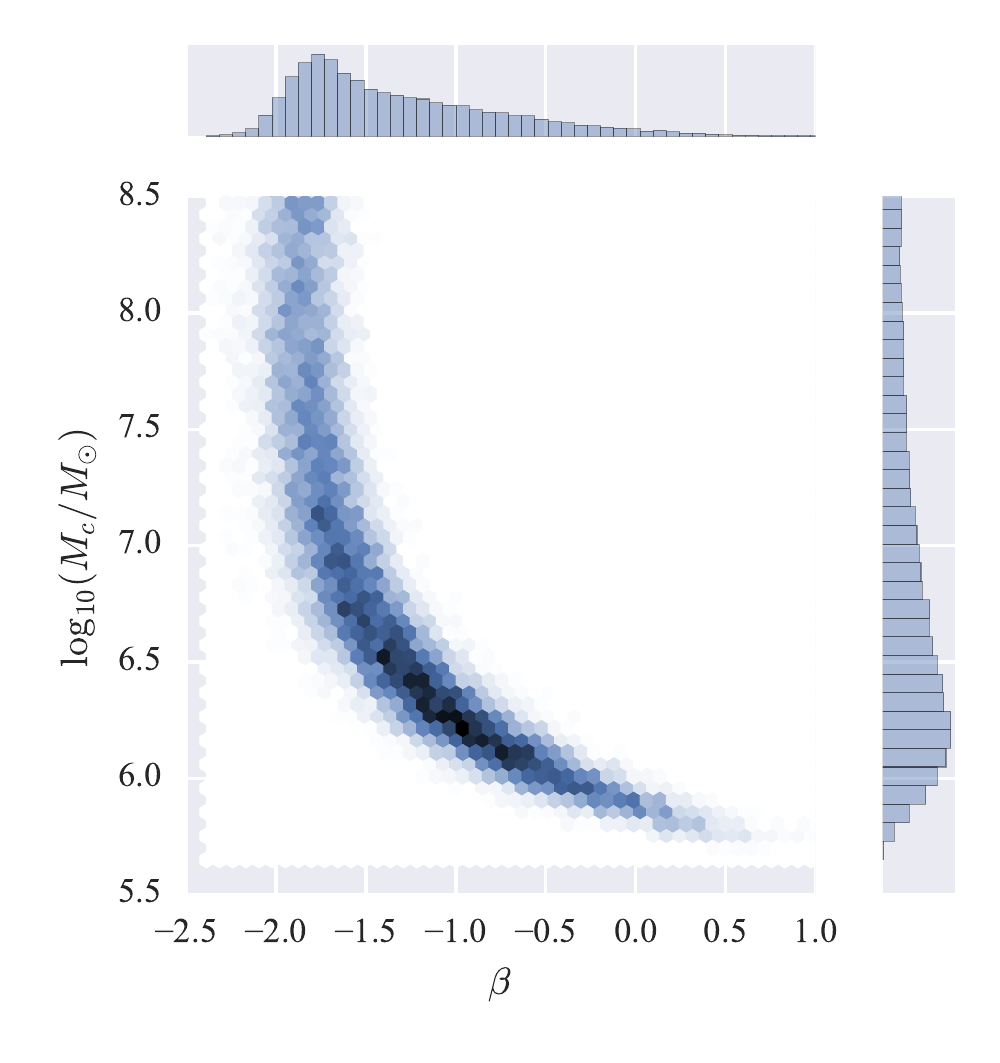}
\caption{\label{fig:sampler} Samples from the posterior probability density function for fits to the truncated power law to the mass distribution of GMCs in the $3.2< R_{g}/\mathrm{kpc}<3.9$ radial bin.  The figure shows the data have some evidence for a cutoff mass at $M_c \sim 2\times 10^{6}~M_{\odot}$, but the data are also well represented with a power law with an index of $\beta = -1.9$.  With either model, the data are consistent with a smaller maximum-mass scale of GMC in this region relative to inner radial bins.}
\end{figure}

In comparing the peak log-likelihoods obtained for the analysis (e.g., $\log_{10} p_{\mathrm{PL}}$) we can determine cases where one model clearly represents the distribution better than the others.  In the inner three radial bins for the analysis, the truncated power-law is clearly preferred over the other two models (see Table \ref{table:properties}).  Indeed, the mass distributions of the clouds are sufficiently shallow that they cannot be fit by a Schechter function at all.  In this case, there is good evidence for a maximum-mass scale and this mass gets smaller with galactocentric radius.  The slopes of the power-law part of the mass distribution are poorly constrained.  However, in the outer three bins, there are relatively few GMCs, precluding good discrimination between these models using empirical distribution functions.  In this case, the three models are not well distinguished though the truncated power law is slightly preferred, but this margin is not larger than the improvement expected by including a second parameter in a fit.  A posterior distribution of parameters for the truncated power-law distribution is shown in Figure \ref{fig:sampler} illustrating the nature of the statistical formulation.  While the truncation mass of $M_c = 2.2 \times 10^{6}~M_{\odot}$ is preferred by the model, the data are also nearly consistent with a pure power-law distribution with $\beta = -1.9$ and a cutoff mass $M_c \gg M_{\mathrm{GMC}}$.  This behaviour typifies the outer three radial bins.

While there is not strong evidence for a specific functional form of the mass distribution in the outer region of the galaxy, we emphasize that all the analysis and data are consistent with a decrease in typical GMC mass in the outer regions of the galaxy.  The index of the pure power-law functional form and, where well constrained, the maximum-mass scale of the clouds $M_c$ decrease with galactocentric radius.  However, with a small number of GMCs in some bins, it is not clear that the truncated power law is a good representation of the distributions and no other functions are identified that are clearly superior.  Thus, we also report non-parametric measurements of the maximum-mass scale, including the maximum-mass cloud ($M_\mathrm{max,GMC}$) as the geometric mean of the five most massive clouds in each bin ($\langle M\rangle_5$).  These estimators also decrease with radius.  We compare this behaviour to that seen in the massive clusters, which also show a decrease in the truncation mass, though the cluster cutoff masses were derived for a Schechter mass distribution.

The truncation mass for the truncated power law, the maximum mass, the typical mass, and the index of the pure power law all decrease with galactocentric radius, with clouds in the centre of M83 having higher typical masses and a shallower mass distribution.  While source confusion may affect nuclear sources, the clouds at $R_{g}>0.5~\mathrm{kpc}$ should be well resolved and the results will be directly comparable to other studies.  Outside of the nuclear region, we see good evidence for a maximum-mass scale for molecular clouds evolving across the face of the galaxy.  This changing mass behaviour is mimicked in the cutoff masses seen in the massive stellar clusters.  Work on cluster mass distributions has suggested a link between cluster mass truncations and the mass distributions of molecular clouds.  Higher mass molecular clouds can host higher mass clusters since the cluster masses are limited to the host cloud mass times the star ($\epsilon$) and bound cluster ($\Gamma$) formation efficiencies.  Since these efficiencies appear to be relatively constant across large areas, the GMC masses should directly limit the cluster masses  \citep{Diederik_Kruijssen_2014}.

The truncation in cluster masses $M_{c,\mathrm{cluster}}$ vary by a factor of 20 whereas the maximum cloud mass scale changes by a factor of 4 ($M_{c,\mathrm{TPL}}$) to 8 ($M_{\mathrm{GMC,max}}$) for $R_g>0.45\mbox{ kpc}$.  The decline in the cluster masses is a product of the declining maximum-mass scales in GMCs and the decreasing cluster formation efficiency \citep{Silva_Villa_2013, Adamo_2015}. There is good correspondence between the maximum cluster mass and the maximum cloud mass, with both decreasing by a factor of $\sim 5$ with radius.  The maximum cluster mass is about 1-2\% of the maximum cloud mass. No data are reported for the clusters for the nuclear region of the galaxy or at radii larger than 4.5 kpc.  However, if the correspondence holds between cluster and cloud masses, we would predict the upper bound on the most massive cluster at $R_g<0.45$~kpc is $4\times 10^5~M_{\odot}$.  This scaling assumes that cluster dissolution is the same in the nuclear region as in the outer regions of the galaxy.  Simulations show that cluster lifetimes are  shorter in the nuclear regions of galaxies \citep{Kruijssen_2011}, which is supported circumstantially by observations \citep{Bastian_2011}.  The most massive cluster in the nuclear region is thus likely to be significantly below this value.

\section{Discussion}
\label{sec:discussion}
\subsection{Mass Scales in M83}
We compare the maximum-mass scales and slopes derived from the empirical distributions to the characteristic masses produced by the Jeans instability and the Toomre instability.  The Jeans instability sets the minimum mass required for a thin sheet of surface mass density $\Sigma$ to overcome random motions with velocity distribution $\sigma_v$.  Such a sheet will fragment into the characteristic (2D) Jeans mass for the system:
\begin{equation}
M_{J} = \frac{\pi\sigma_v^4}{4G^2\Sigma}.
\end{equation}
By contrast, a thin shearing disk with rotation curve $V(R_g)$ as a function of galactocentric radius $R_g$ will fragment into a characteristic mass set by the \citet{Toomre_1964} instability, below which self-gravity can overcome shear.  The largest unstable scale of the Toomre instability is $\lambda_T = 4\pi^2 G \Sigma \varkappa^{-2}$ where $\varkappa$ is the epicyclic frequency, leading to a characteristic mass of 
\begin{equation}
M_T = \frac{\pi}{4} \lambda_T^2 \Sigma = 4\pi^5\frac{ G^2\Sigma^3}{\varkappa^4}.
\end{equation}
We infer $\Sigma$, $\sigma_v$ and $\varkappa$ from the ALMA data and supplementary information.  We include the {\sc Hi} 21-cm map of the galaxy that is part of the THINGS \citep{Walter_2008} survey in our analysis.  The low-resolution studies of \citet[][L04a]{Lundgren_2004} and \cite[][L04b]{Lundgren_2004b} mapped the galaxy in CO$(1\to0)$ and CO$(2\to1)$ emission using SEST achieving a resolution of $27''$ at best.  While low resolution, this work provides a good measure of the overall gas masses without the effects of spatial filtering.

\begin{figure}
\begin{center}
\includegraphics[width=\columnwidth]{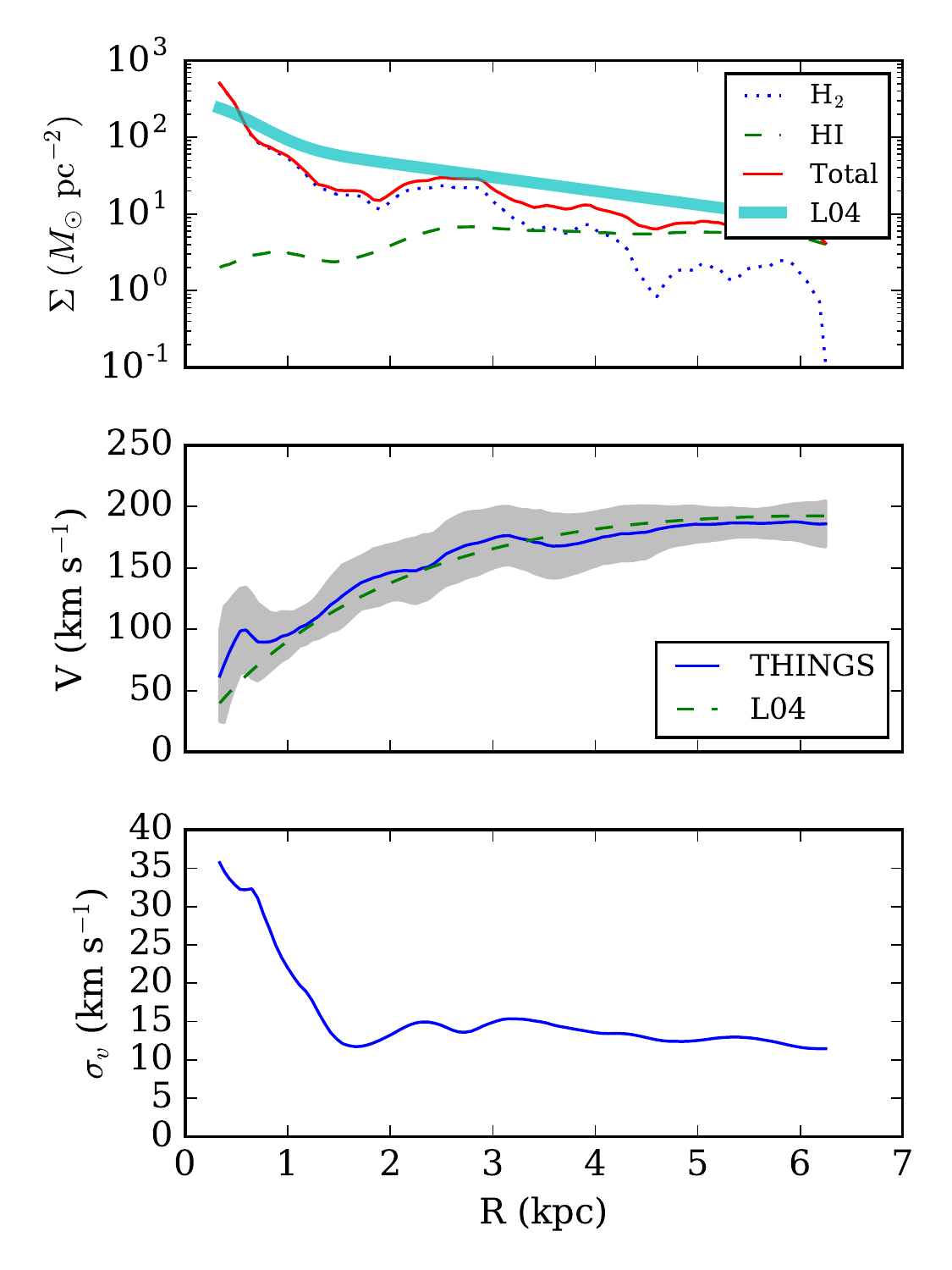}
\caption{\label{fig:profiles} Profiles used to derive characteristics masses for the M83 disk. (top) Surface density profiles obtained from averaging the ALMA CO brightness and the THINGS 21-cm brightness in radial bins.  The L04ab work is shown for comparison. (middle) This panel shows the rotation curve of the \citet{Lundgren_2004} study of M83 using low resolution CO mapping (dashed line).  Using the 21-cm THINGS data of \citet{Walter_2008}, the figure indicates the median inferred rotation speed (solid curve) and absolute deviation around that median (shaded region) for the atomic gas.  The curves show good agreement.  (bottom) ISM velocity dispersion inferred from the ALMA CO and THINGS 21-cm data.  The velocity dispersion is high in the centre of the galaxy but approaches a constant once outside of the central region.%
}
\end{center}
\end{figure}

We measure ISM properties from the THINGS and ALMA data cubes, using the average line-of-sight velocity measurements of the 21-cm data to provide a constraint on the velocities at which the neutral ISM will be found in any given spectrum.  We then shift those assumed velocities to average spectra in radial bins, thereby determining an average line profile, even when the line cannot be readily discerned or characterized \citep[see][for details]{Schruba_2011}.  To measure the surface density profiles, we shift and average the profiles in radial bins and convert the integrated spectra to surface densities.  The profiles of the surface density curves are shown in Figure \ref{fig:profiles}.  There is reasonably good agreement between the ALMA data and the L04a work, after scaling to our adopted CO-to-H$_2$ conversion factor.

Measuring $\varkappa$ and $\sigma_v$ for the gas disk requires measuring the rotation curve.  The analysis of L04b uses the low resolution CO mapping to derive a rotation curve.  They find the rotation curve is well modeled by an exponential disk.  We confirm this by using their kinematic parameters (i.e., inclination and position angle) to estimate the amplitudes of the rotational motion for the atomic gas.  In Figure \ref{fig:profiles}, we show the L04b rotation curve and the median absolute deviation of inferred rotational velocities for the 21-cm data around the rotation curve (grey region).   There is good agreement within the scatter between the two approaches.  We can derive the value of $\varkappa$ using the smooth functional form of the L04b curve:
\begin{equation}
V^2 = \frac{2 G M_d}{R_d} y^2 [I_0(y)K_0(y)-I_1(y)K_1(y)]
\end{equation}
where $y\equiv R_g/(2R_d)$, $M_d$, the disk mass $M_d=6.4\times 10^{10}~M_{\odot}$, the scale length $R_d=2.9$~kpc, and $I(y),K(y)$ are modified Bessel functions.  For this curve, the epicyclic frequency is 
\begin{equation}
\varkappa^2 = \frac{G M_d}{R_d^3} \left\{ \left[2 I_0(y)+yI_1(y)\right]K_0(y)-\left[y I_0(y)+I_1(y)\right]K_1(y)\right\}.
\end{equation}
Using a numerical derivative and the THINGS rotation curves decreases the average Toomre mass by a factor of 8 inside the nucleus, but the results are otherwise comparable.  Since the resolution of the THINGS data is better than the L04a analysis ($15''$ vs $27''$) we use those data in what follows.  Finally, we measure the ISM velocity dispersion using the {\sc Hi} and CO data.  For each radial bin, we shift the spectra to a common centre velocity and measure the intensity-weighted second moment of the resulting profile.  The resulting velocity dispersion is shown in the bottom panel of Figure \ref{fig:profiles}.  We use the mass-weighted sum in quadratures of the two velocity dispersions.  In the nuclear region, this velocity dispersion is likely an upper limit owing to beam smearing within the 300 pc beam {\sc Hi} beam and the velocity centroid not representing the circular velocity of emission properly. Outside of the nuclear region ($R_{\mathrm{gal}}>1\mbox{ kpc}$), these effects become small ($<5\mbox{ km s}^{-1}$).  However, inside this region, the velocity dispersion will be inflated by these errors and will require higher resolution observations and a more careful analysis to measure properly.

We measure the surface density of gas from the Toomre criterion using the THINGS 21-cm data combined with either the L04a data or the ALMA observations.  We report these as $M_{\mathrm{Toomre}}$ (THINGS+L04)and $M^*_{\mathrm{Toomre}}$ (THINGS+ALMA) in Table \ref{tab:masses}.  Since the L04a CO observations measure all the emission at low resolution, they capture the total molecular mass in the galaxy disk.  The ALMA data measure the mass on scales smaller than 200 pc, corresponding to the largest angular scale recovered in these observations.  The relevant surface densities to consider are the masses up to the scales of the Toomre instability ($\lambda_T\sim 500$~pc), which are not completely sampled in the interferometer data.  The single dish data likely measures both low mass clouds below our completeness limit as well as diffuse emission unconnected to the Toomre mechanism, such as the large scale-height gas seen in \citet{Pety_2013}, as well as fainter emission that is concentrated near the molecular clouds, which should be considered in the estimate.  Thus, the Toomre mass should be considered to be larger than (but comparable to) the ALMA-based values ($ M^*_{\mathrm{Toomre}}$) with a maximum scale given by the single dish data ($ M_{\mathrm{Toomre}}$).  The inner region has $M^*_{\mathrm{Toomre}}>M_{\mathrm{Toomre}}$ because the coarse resolution of the single dish data is unable to resolve the sharp peak in the molecular gas surface density in the nucleus.

\begin{table*} 
    \begin{tabular}{ c r r r r r r}
\hline
    Bin     & $M_{\mathrm{Jeans}}$ & $M_{\mathrm{Toomre}}$ &$M^*_{\mathrm{Toomre}}$ & $\langle \Sigma_{\mathrm{H2}}\rangle$ & $\langle P_{\mathrm{int}}/k_\mathrm{B} \rangle$ & $\langle t_{\mathrm{ff}} \rangle$\\
    (kpc)   & ($10^6~M_{\odot}$) & ($10^6~M_{\odot}$) & ($10^6~M_{\odot}$) &($M_{\odot}$ pc$^{-2}$) & ($10^5$ K cm$^{-3}$) & (Myr)\\ 
    \hline
    0 - 0.45   & 60 &  14 & 94 & 700 & 68 & 2\\
    0.45 - 2.3 &  60 & 31 & 14&  180 & 9 & 5\\
    2.3 - 3.2  &  20 & 15 & 3.7&  150 & 4 & 5\\
    3.2 - 3.9  &  40 & 46 & 3.3&  160 & 4 & 5\\
    3.9 - 4.5  &  60 & 14 & 1.1&  120 & 3 & 5\\
    $>4.5$     &  90 & 16 & 1.6&  150 & 3 & 5\\
\hline
    \end{tabular}
    \caption{\label{tab:masses} Theoretical masses and cloud properties and the average internal conditions for GMCs, including the surface densities ($\Sigma_{\mathrm{H2}}$), the internal pressures ($P_{\mathrm{int}}$), and the free-fall times ($t_{\mathrm{ff}}$).  The two different measurements for the Toomre mass refer to calculating the molecular gas surface density profile from either the single dish data of L04a ($M_{\mathrm{Toomre}}$) or the ALMA data ($M^*_{\mathrm{Toomre}}$).  These mass scales are calculated from the radial profiles and averaged with mass-weighting.}
\end{table*}

Table \ref{tab:masses} shows the variation in the characteristic masses for the ISM in the system using the radial profiles described above.   Both the Jeans mass and the Toomre mass are larger than the maximum cloud mass across the disk.  Formally, the Toomre mass represents an upper limit to the cloud masses, set by rotational shear and the Jeans mass represents a typical mass for fragmentation.  Both are significantly larger than the structure we see in the disk.  Neither of the theoretical masses show significant trends with radius apart from a modest increase in the nuclear region of the galaxy.  The Toomre masses in the inner regions of the galaxy are likely overestimates since the L04b rotation curve and low-resolution THINGS data we use here neglect the complex dynamical environment in the inner 500 pc of M83 associated with its double nucleus  \citep{Sakamoto_2004,Rodrigues_2009}.  Furthermore the shear rate in the bar will be larger than predicted from the circular velocity curve alone owing to the substantial non-circular motions.  In these two regions, the Toomre mass limits will be overestimates of the fragmentation scale in the region.

In general, we find that $M_{\mathrm{Jeans}} > M_{\mathrm{Toomre}}$. Because the former is a minimum mass scale and the latter is a maximum-mass scale, this result would formally indicate that collapse cannot take place.  In the above definitions, the case $M_{\mathrm{Jeans}} = M_{\mathrm{Toomre}}$ refers to Toomre $Q=2$, which implies marginal stability. With $M_{\mathrm{Jeans}} > M_{\mathrm{Toomre}}$, we have $Q>2$ across the disk, so it is stable to collapse, except possibly in the centre though the dynamical environment is not well probed in that regime. In the stable disk case, the Toomre mass may still be the guiding factor for cloud collapse: because turbulence gets dissipated on small scales, the local Jeans mass will drop, implying that the first fragmentation takes place at the Toomre mass.  If the relevant physical limits are closer to the interferometer-only surface densities, $M^*_{\mathrm{Toomre}}$, the masses would agree well with the maximum-mass scales that we see in the mass distribution.  Assessing the stability rigorously requires (now in-progress) ALMA observations that recover the full range of spatial scales.  However, either estimate of the Toomre mass points to this shear regulation being an important factor in shaping the cloud mass distribution \footnote{We note that shear may not regulate cloud masses in all environments. Under the low-shear and low-surface density conditions of outer galaxy discs, feedback has been suggested to set the high-mass end of the cloud mass function, resulting in maximum cloud masses lower than the Toomre mass \citep{reinacampos17}}.

The maximum-mass scales of the clusters should be linked to the characteristic masses in the ISM \citep{Diederik_Kruijssen_2014} with the cluster mass given as:
\begin{equation}
M_{c,\mathrm{cluster}} = \epsilon \Gamma M_{c,\mathrm{GMC}},
\end{equation}
where $\epsilon$ is the star formation efficiency (dimensionless) and $\Gamma$ is the bound cluster formation efficiency.  The cluster formation efficiency has been observed in M83 to decrease with radius \citep{Silva_Villa_2013, Adamo_2015} with typical values of $\sim$10\%.   Star formation efficiencies of molecular clouds are $\sim$10\% \citep{Lada_2003}. Combining these two efficiencies suggests the observed values of $M_{c,\mathrm{cluster}}/ M_{c,\mathrm{GMC}} = 10^{-2}$ is largely consistent with expectations.  For $R_g>0.5$~kpc, the observed cluster formation efficiency $\Gamma$ drops by a factor $\sim 5$ over the region studied \citep{Silva_Villa_2013, Adamo_2015}, but the internal conditions of the molecular clouds do not change significantly in the disk of the galaxy.  The clouds all show roughly constant properties including turbulence, surface density, internal pressure and free-fall time (Figure \ref{fig:radial}, Table \ref{tab:masses}).  This result is consistent with the view that the cluster formation efficiency is set by the gas surface density on large scales rather than the internal cloud conditions \citep{Kruijssen_2012}.  

\subsection{Systematic Effects}
\label{sec:systematics}
The above analysis points to a good connection between cluster mass and cloud mass in M83.  However, two systematic effects can potentially alter these results.  Here we review the nature of the ALMA data and the effects of the cataloging algorithm to assess the nature of these uncertainties.

{\it Data Quality} -- The quality of ALMA data vastly exceeds nearly every other previous study of extragalactic molecular clouds, and the sensitivity and resolution of this study are excellent compared to the preceding work that frames this study.  However, these results only use data drawn from the 12-m dishes on ALMA.  As such, they are affected by spatial filtering of emission.  The comparison of the surface density profiles to those of L04a highlights that ALMA recovers between 50\% and 100\% at every radius.  Furthermore, the data cube shows negative sidelobes around some of the bright sources, particularly in the centre of the galaxy.  Such effects are typical in extragalactic studies \citep{Rosolowsky_2006}.  While this study does not include single dish data, the work of \citet{Pety_2013} studying M51 demonstrated that most the filtered emission was associated with a diffuse, high-line-width emission component and rather than high-mass molecular clouds.  By analogy, we suspect that most of the emission that is being filtered out is likely to be a diffuse CO($1\to0$) emission component, with ALMA recovering the bright, compact structures in the data, namely the high-mass molecular clouds.  However, such ``diffuse gas'' is indistinguishable from a population of low mass clouds below our completeness limit.  Our mass functions reported in Table \ref{table:properties} are poorly constrained at the low mass end, but extrapolating the pure power-law mass distributions from the completeness limit down to $M=0$ could account for most of the difference between the cataloged masses and the total single-dish mass observed by L04a.  The nature of the diffuse or low-surface brightness emission is not well described in this data though forthcoming work with ALMA will help clarify the nature of this emission in this target.  The excellent agreement of the cloud properties seen here with previous studies suggests that the analysis is doing an adequate job of recovering the molecular clouds above the sensitivity limit and characterizing their properties.  The maximum recovered scale for these data should be $\sim 10\times$ the beam scale or 200 pc, and the velocity gradient of the galaxy keeps emission in each channel confined spatially.  Thus, we should get a relatively good measurement of cloud properties, of quality comparable to or exceeding previous work.

{\it Cloud Decomposition} -- The major systematic uncertainty in the analysis is the cloud decomposition algorithm.  {\sc cprops} is designed to provide a stable decomposition of ISM structure in the low signal-to-noise case, accounting for the effects of interferometers.  However, the algorithm does not have a large dynamic range in the scales of objects that it recovers.  We have set the parameters to stabilize the cloud recovery on scales comparable to the prior cloud structures expected from Milky Way studies.  Since the algorithm attempts to assign all emission in the data cube to molecular clouds that are well separated, the analysis will tend to join low mass clouds to neighbouring high mass clouds.  The combined object will be seen as a single, high mass object.  However, if the underlying mass distribution is steep (say a power law distribution with $\beta \sim -2$) and clouds are drawn randomly from the mass distribution, the high mass cloud will tend to be significantly more massive than the low mass cloud.  The effect of such combination is to remove clouds from the low mass portion of the mass spectrum and add them to the high mass clouds.  This will make the mass distribution appear artificially shallow at the low mass end.  Correctly separating the low mass clouds from their high mass neighbours would create an even sharper truncation in the distribution and it is unlikely to move significantly in value.  Higher resolution and sensitivity observations may serve to better constrain the index of the mass distribution near the mass limit of our study ($M\gtrsim 3.35\times 10^{5}~M_{\odot}$).

The blending effects will be particularly acute in the galaxy centre where the separation between the clouds is comparable to the cloud size.  We should therefore regard the clouds in the $R_g<0.45$~kpc bin as particularly suspect and not necessarily representing distinct physical entities.  The average characteristics of the ISM in this area are still a useful measure of the changing internal conditions of the clouds.  Thus, the mass distribution is suspect, but the ISM clearly has higher turbulent velocity dispersions, average densities, but not significantly different degrees of gravitational binding.  However, in the outer disk of the galaxy, the clouds should be well separated and the mass distributions are closer to the true distributions.  It is in this outer region where we can make a clearer association between the cloud masses and cluster masses, finding a link consistent with theoretical expectations.

\section{Conclusions}

We present a cloud-based analysis of the molecular gas in M83 as observed by ALMA.  We compare the results of the cloud decomposition to the properties of the young massive cluster population, searching for a connection between the structural organization of the molecular gas and the changing cluster properties.  Based on this analysis, we reach the following conclusions:
\begin{enumerate}
\item The molecular clouds in M83 are well-resolved in the ALMA data and show excellent correspondence with scaling relations seen in other systems.  On average, they are consistent with significant self-gravitation and a turbulence driven size-line width relationship.
\item Despite the overall correspondence between the molecular cloud populations and the scalings seen in other systems, there are systematic variations in cloud properties over the face of the galaxy.  Of note, the clouds found in the nuclear region ($R_g<0.5$~kpc) have significantly higher surface densities ($\langle \Sigma \rangle = 700~M_{\odot}~\mbox{pc}^{-2}$ vs. $170 ~M_{\odot}~\mbox{pc}^{-2}$ in the disk) and turbulent line widths on 1 pc scales $\langle \sigma_0\rangle = \langle \sigma_v R_0^{-0.5}\rangle = 1.7 \mbox{ km s}^{-1}$ vs. $0.7 \mbox{ km s}^{-1}$ in the disk.  The higher densities and more intense turbulence of the central clouds balance so that these clouds have gravitational binding energies comparable to their internal kinetic energies.  This result is shown by virial-theorem-based estimates for cloud mass being consistent with mass estimates from their CO luminosity (i.e., the $X$-factor).  These differences are found to be consistent with theoretical expectations for clouds in a higher surface density environment.
\item The mass distributions of molecular clouds change over the face of the galaxy.  There is good evidence for a maximum-mass scale in the population, which sets an upper limit for molecular cloud mass.  Functional fits to the mass distribution are consistent with this conclusion but there is not strong evidence for a particular functional form in the outer radial bins of our analysis.  The maximum mass in the population is highest in the centre of the galaxy though blending of emission features likely biases this result.  Outside of the nucleus, the maximum-mass cloud found in bins of equal area decreases by a factor of 8.  The behaviour of the cloud mass distribution at the low-mass end is poorly constrained, likely because of blending, which is most problematic in the central region.
\item Truncation masses have been previously observed in the cluster population and fit with a Schechter function, namely a power-law mass distribution with and index of $\beta=-2$ and an exponential cutoff above a truncation mass.  There is not good evidence for this being the best representation of the molecular cloud mass distribution.  The maximum molecular cloud mass in a bin is $\sim 10^2$ times the maximum cluster mass.
\item Except in the galaxy centre, we find the maximum cloud masses are comparable to the predictions from the Toomre criterion, which is the mass scale on which structures will form in a shearing disk.  The disk appears to be globally stable with respect to gravitational collapse, so that local shear is likely a primary regulator of cloud mass.  This conclusion will depend on the detailed distribution of matter on the scales of the Toomre instability ($>200$ pc), which are not measured in the ALMA observations.
\item The maximum-mass cluster is $1-2\%$ of the mass of the maximum-mass molecular cloud, which is consistent with a simple correspondence model where clouds form stars with a dimensionless efficiency of $10\%$ and the observed cluster formation efficiency (i.e., the fraction of formed stars that remain in bound clusters) being the observed $\Gamma\sim10\%$.  The cluster formation efficiency is observed to vary with radius over the face of the galaxy by a factor of a few but, for $R_g>0.5\mbox{ kpc}$, the internal conditions of the molecular cloud population remain nearly constant.  This result is consistent with theoretical predictions given the observed gradient in the gas surface density and pressure.
\end{enumerate}

Future work, particularly in-progress observations with ALMA, will be able to extend this type of analysis over the entire disk of M83 using data including all spatial frequencies.  In particular higher resolution observations of the nucleus will highlight the evolving properties of molecular clouds in this region and make an unambiguous measurement of the mass distribution.

\medskip
%\acknowledgements

We are grateful for the comments of an anonymous referee, who encouraged a more careful treatment of the statistics in the manuscript.  This paper makes use of the following ALMA data: ADS/JAO.ALMA\#2012.1.00762.S. ALMA is a partnership of ESO (representing its member states), NSF (USA) and NINS (Japan), together with NRC (Canada) , NSC and ASIAA (Taiwan), and KASI (Republic of Korea), in cooperation with the Republic of Chile. The Joint ALMA Observatory is operated by ESO, AUI/NRAO and NAOJ.  The National Radio Astronomy Observatory is a facility of the National Science Foundation operated under cooperative agreement by Associated Universities, Inc.  The data were retrieved from the JVO portal (\url{http://jvo.nao.ac.jp/portal}) operated by the NAOJ.  This work made use of the Astropy package for Python \citep{Astropy} and we gratefully acknowledge the ongoing efforts of our community's developers.  This work also made use of the NASA Astrophysics Data System and the NASA Extragalactic Database.  PF and EWR are supported by a Discovery Grant from NSERC of Canada. JMDK gratefully acknowledges financial support in the form of a Gliese Fellowship and an Emmy Noether Research Group from the Deutsche Forschungsgemeinschaft (DFG), grant number KR4801/1-1.  NB is partially funded by a Royal Society University Research Fellowship and an European Research Council (ERC) Consolidator Grant (Multi-Pop - 646928).

\bibliography{ms.bib}
\label{lastpage}

\end{document}